\documentclass{elsart}

\usepackage{natbib}

\usepackage{graphicx}

\usepackage{amssymb}

\begin{document}

\begin{frontmatter}

\title{Chemical etching of a disordered solid: from experiments to
field theory.}

\author[ad1]{Andrea Gabrielli}
\address[ad1]{SMC-INFM and ISC-CNR, Dipartimento di
Fisica, Universit\`a ``La Sapienza'' di Roma, P.le Aldo Moro 2, 0185 -
Rome, Italy.}
\author[ad2]{Andrea Baldassarri}
\address[ad2]{INFM, Dipartimento di
Fisica, Universit\`a ``La Sapienza'' di Roma, P.le Aldo Moro 2, 0185 -
Rome, Italy.}
\author[ad3]{Miguel Angel Mu\~noz}
\address[ad3]{Institute ``Carlos I'' for Theoretical and Computational 
Physics, University of Granada, 18071-Granada, Spain.}
\author[ad4]{Bernard Sapoval}
\address[ad4]{Laboratoire de la Physique de la Mati\`ere Condens\'{e}e, Ecole
Polytechnique - CNRS, 91128 - Palaiseau, France.}

\begin{abstract}
We present a two-dimensional theoretical model for the slow chemical
corrosion of a thin film of a disordered solid by suitable etching
solutions. This model explain different experimental results showing
that the corrosion stops spontaneously in a situation in which the
concentration of the etchant is still finite while the corrosion
surface develops clear fractal features. We show that these properties
are strictly related to the percolation theory, and in particular to
its behavior around the critical point. This task is accomplished both
by a direct analysis in terms of a self-organized version of the
Gradient Percolation model and by field theoretical arguments.
\end{abstract}

\begin{keyword}
Disordered solid \sep corrosion \sep gradient percolation \sep field theory
\sep absorbing state phase transitions
\end{keyword}

\end{frontmatter}

\section{Introduction}
\label{intro}
When an etching solution is put in contact with a disordered etchable
solid, it corrodes the ``weak'' parts of the solid surface while the
``hard'' parts remain un-corroded. During this process new regions of
the solid (both hard and weak) are discovered coming into contact with
the etching solution.  If the volume of the solution is finite and the
etchant is consumed in the chemical reaction, the etching power of the
solution diminishes progressively and the corrosion rate slows down.
When the solution is too weak to corrode any part of the hardened
solid surface, the dynamics spontaneously stops.  It is an
experimental observation \cite{Balazs} that the etchant concentration
at the arrest time is larger than zero. We show below that its value
is strictly related to the percolation threshold of the considered
solid lattice. We show also that the final connected solid surface has
clear fractal features up to a certain characteristic scale $\sigma$,
i.e., the surface thickness.  This is precisely the phenomenology that
has been recently observed in experiments on pit corrosion of aluminum
thin films \cite{Balazs}.  We show that the fractal features and the
characteristic scale can be explained by the critical behavior of
classical percolation around the percolation threshold \cite{Aharony}.

\section{The model}

A simple dynamical model, capturing the above mentioned
phenomenology, has been recently proposed and studied
\cite{model,GBS}.
\begin{figure}
\centerline{\includegraphics[width=0.4\textwidth,height=0.5\textheight,
angle=-90]{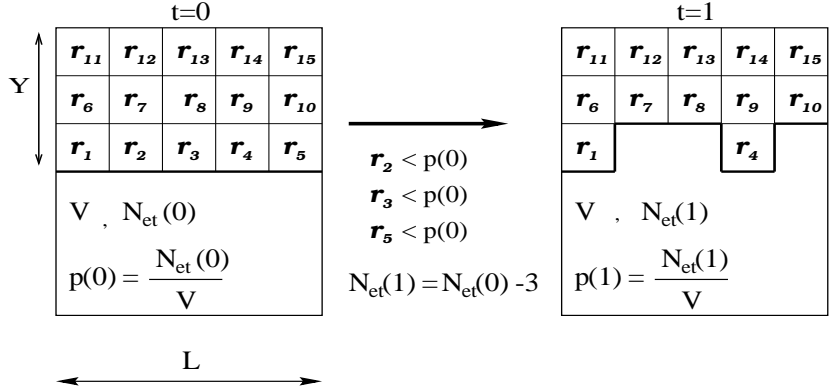}} 
\caption{Sketch of the etching dynamics in a square lattice: the sites
 $2,3,5$ are etched at the first time-step as their resistances are
 lower than $p(0)$.  At the same time the number of etchant particles
 in the solution decreases by $3$ units, and a new part of the solid
 is uncovered.
\label{fig1}}
\end{figure} 
The model is sketched in Fig.~\ref{fig1}, and can be formulated as
follows:\\ (i) The solid is represented by a square lattice of sites
with random resistances to corrosion $r_{i}$ uniformly distributed in
the interval $[0,1]$.  It has a width $L$ and a given fixed depth $Y$.
The etching solution has a finite constant volume $V$, and contains an
initial number $N_{et}(0)$ of etchant molecules: the initial etchant
concentration is therefore $C(0)=N_{et}(0)/V$.  Experimentally the
``etching power'' $p(t)$ of the solution at time $t$ is roughly
proportional to the concentration $C(t)$. We assume $p(t)=C(t)$
without loss of generality, and take $p_c<p(0)\le 1$, where $p_c$ is
the percolation threshold of the lattice.\\ (ii) At $t=0$ the
solution is put in contact with the solid through the bottom boundary
$y=0$.  At each time-step $t$, the solid sites belonging to the
solid-solution interface with $r_{i}<p(t)$ are removed, and a particle
of etchant is consumed for each corroded site.  Consequently, the
concentration of the solution decreases with $t$.  Finally, depending
on the connectivity criterion chosen for the lattice sites (e.g., first 
nearest neighbor connectivity for solid sites), $m(t)$ new
solid sites, previously in the bulk, come into contact with the
solution for the first time.  At the next time-step, only
these sites can be corroded, as the other surface sites have
already resisted to etching and $p(t)$ decreases with $t$.  Calling
$n(t)$ the number of removed solid sites at time $t$, and $N(t)$ the
total number of removed sites up to time $t$, one can write
\begin{equation}
p(t+1)=p(t)-\frac{n(t)}{V}=p(0)-\frac{\sum_{t'=0}^t
n(t)}{V}=p(0)-\frac{N(t)}{V}\,.
\label{p-n}
\end{equation}
Note that, during the process, the solution can surround and detach
finite solid islands from the bulk.  The global solid
surface is then composed by the union of the perimeter of the ``infinite''
solid, here called {\em corrosion front}, and the surfaces of the
finite islands. At the end, the corrosion dynamics spontaneously stops
at $t_f$ such that all the surface sites have resistances $r>p(t_f)$.

The main features of the model, found through extensive numerical
simulations (lattices with up to $2000\times 2000$ solid sites in 
\cite{GBS}), are:\\ 
(i) The final value $p_f=p(t_f)$ is slightly
smaller than the percolation threshold $p_c$.  The difference
$|p_f-p_c|\rightarrow 0$ as $(L/V)^{\alpha_p}$ (we take $L/V \ll 1$)
with $\alpha_p\simeq 0.45$, when the limits $L,V\rightarrow +\infty$
are taken in the appropriate way \cite{GBS}.
All this can be explained through percolation theory which, in fact,
implies that for $p(t)\le p_c$ the probability to have a connected
path of solid sites all with $r>p(t)$ (and then stopping
corrosion) is equal to one in the infinite volume limit.
\begin{figure}
\centerline{\includegraphics*[width=\textwidth]{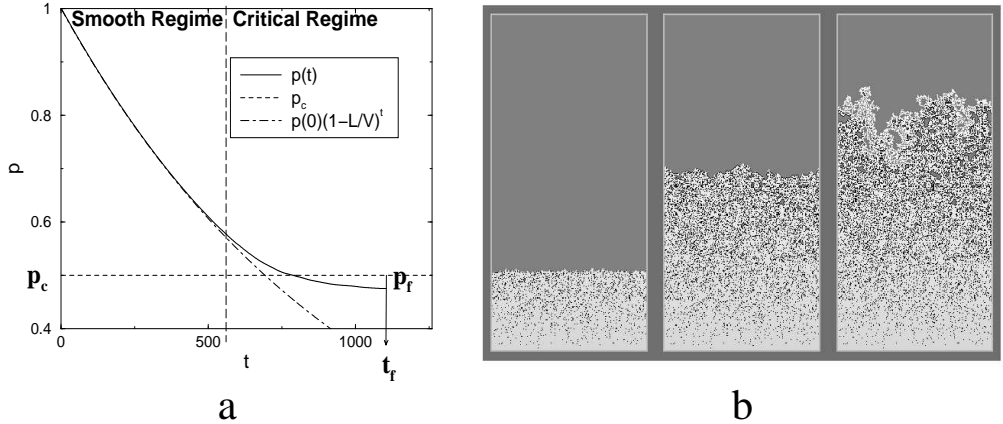}}
\caption{{\bf a}- Time decay of $p(t)$ (continuous line), compared to
the analytical approximation $p(t)=p_0\exp (-tL/V)$ in the smooth
regime (dashed line) \cite{GBS} in a triangular lattice.  There is
good agreement for $p(t)>p_c$.  {\bf b}- Three snapshots of the
corroded solid (dark gray) by the solution (clear) at three different
time-steps: initial, intermediate and final.  Black regions represent
detached islands.  The final corrosion front shows fractality up to a
scale given by its thickness.
\label{fig2}}
\end{figure}
\\(ii) The corrosion dynamics can be divided (see Fig.~\ref{fig2}-{\bf
a}) into two regimes: a {\em smooth} regime when $p(t)>p_c$, and a
{\em critical} regime when $p(t)\simeq p_c$.  The duration of the
former is measured approximately by $t_c$, defined by $p(t_c)=p_c$;
$t_c$ is found to scale with the ratio $L/V$ in the following way
$t_c\sim V/L$.  The duration of the latter, $(t_f-t_c)$, scales as
$(V/L)^{\alpha_{t_f}}$ with $\alpha_{t_f}\simeq 0.55$ (with a further
linear dependence on $\log\,L$ of the scaling coefficient due to the
extremal nature of $t_f$ \cite{frac}).\\ (iii) The surface of the
solid in contact with the solution displays a peculiar roughening
dynamics (see Fig.~\ref{fig2}-{\bf b}).  In the first smooth regime,
the corrosion has a clear mean direction given by the initial
condition, and the corrosion front becomes progressively rougher and
rougher, while finite detached islands are quite small.  In the second
critical regime, spatial correlations increase on the corrosion front and 
the dynamics generates a locally isotropic fractal
geometry, while the detached islands are larger.
\begin{figure} 
\centerline{\includegraphics[width=0.9\textwidth]{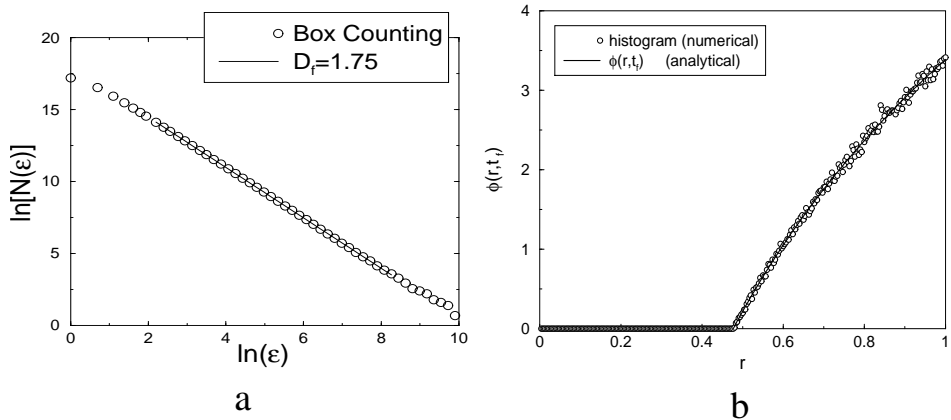}} 
\caption{{\bf a}-Box counting measure of the fractal dimension
of the corrosion front for length smaller than $\sigma$. {\bf b}-Final
normalized histogram of the global solid surface: simulations results
(empty points) are compared with the analytical evaluation (continuous
line).  
\label{fig3}} 
\end{figure} \\
(iv) The final corrosion front is fractal with a fractal dimension
$D_f=1.753\pm 0.05$ (see Fig.~\ref{fig3}-{\bf a}) up to a
characteristic scale $\sigma$ which is the final thickness of the
front.  On larger scales the corrosion front looks like a
one-dimensional line.  Also $\sigma$ scales with $L/V$ as follows:
$\sigma\sim (L/V)^{-\alpha_{\sigma}}$ with $\alpha_{\sigma}\simeq
0.57\simeq 1/D_f$.  All these properties are well explained in the
framework of percolation theory as a self-organized version of the
Gradient Percolation model \cite{GP,GBS}.\\ (v) Another important
experimental observation captured by this simple model is the
progressive hardening of the solid surface, due to the corrosion only
of freshly discovered weak sites which, in turn, deplete the etchant
concentration in time permitting future corrosion of only weaker
sites.  This hardening effect is described quantitatively by the
behavior of the normalized resistance histogram $\phi(r,t)$ of the
global surface sites giving the density of such sites with resistance
$r$.  Obviously, $\phi(r,0)=1$ with $r\in [0,1]$, while $\phi(r,t_f)$
is given in Fig.~\ref{fig3}-{\bf b}. As one can see, the final global
surface is much more resistant to corrosion than the initial one.  The
time evolution and the final shape of $\phi(r,t)$ have been
successfully obtained theoretically in \cite{GBS}.

\section{Field theory approach and dynamical percolation}

In order to describe this model in the critical regime, it is possible
to develop a phenomenological field theory approach ``\`a la Landau''
(see \cite{Munoz}) consisting in writing down the functional Langevin
equation of the process around criticality directly from the analysis of
the symmetries of the system.  To this aim let us consider the following three
different local densities or coarse-grained fields:\\ (i) $s({\bf
x},t)$ describing the local density in the point ${\bf x}$ at time $t$
of material susceptible to be etched at any time after $t$ (i.e., in
the discrete model presented above, bulk solid sites and ``fresh''
solid surface sites freshly arrived to the solid-liquid interface).\\
(ii) $q({\bf x},t)$ is the local density of passivated and inert
material (i.e., surface solid sites having already resisted an etching
trial and then immune or not-susceptible to be corroded at any future
time-step). \\ (iii) $c({\bf x},t)$ is the local density of corroded
sites replaced by solution sites.

The mean field equations (rate equations) describing the evolution of
the averaged mean values of these magnitudes are to the leading order
in the fields \cite{Munoz}:
\begin{eqnarray}
\dot{s}(t) & = & -  \alpha         c(t) s(t) \nonumber  \\
\dot{q}(t) & = &    \alpha (1-p(t)) c(t) s(t)  \nonumber \\
\dot{c}(t) & = &    \alpha  p(t)    c(t) s(t)  
\label{deterministic} 
\end{eqnarray}
where $p(t)$ is the probability to etch an active site at time $t$,
and $\alpha$ is a positive constant that we fix equal to one without
loss of generality. The interpretation of the first equation is: in
order for the density of susceptible sites to change (decrease) in a
region, it is necessary to have locally both a non-vanishing density
of etchant and raw solid material susceptible to be etched.  This
restricts the dynamics to {\it active} regions (i.e., part of the
solution-solid interface) in which non-vanishing local densities of
$s$ and of $c$ coexist.  The second and the third relations in
Eq.~\ref{deterministic} say that an active site becomes either a
$c$-site, with probability $p(t) $, or alternatively, after healing, a
$q$-site with complementary probability $1-p(t)$.  Note that, as
$\dot{s}~+~\dot{c}~+~\dot{q}~=~0$, the total number of sites is
conserved during the dynamics. We have written so far mean field
equations in which spatial dependence and fluctuations are not taken
under consideration.  To this aim, it is convenient to introduce the
activity field $\rho({\bf x},t)\equiv c({\bf x},t)s({\bf x},t)$.
From Eq.~\ref{deterministic} it follows immediately that
\begin{equation} 
\dot{\rho}(t) = - c(t) \rho(t) + p(t) s(t) \rho(t) \,.
\label{det2}
\end{equation}
We now use for $p(t)$ Eq.~\ref{p-n}, noting that in terms of the
coarse grained fields we have $N(t)=\int d{\bf x} [c({\bf x},t)-c({\bf
x},0)]$.  Using this observation, it is simple to derive the
expression of $p(t)$ as a function of the activity field $\rho({\bf
x},t)$:
\begin{equation}
p(t)=p(0) \exp\left[ - {1 \over V} \int_0^t dt'
 \int_V d{\bf x} \rho({\bf x},t') \right]=
p(0) \exp\left[-\int_0^t dt' \overline{\rho}(t') \right]\,,
\label{p3}
\end{equation}
where $\overline{\rho}(t)={1 \over V}\int d{\bf x} \rho({\bf x},t)$ is
the average activity at time $t$.  Since $p(t)$ is a function only of
the average activity, it can be considered as a deterministic, smooth,
positive and decreasing term in the final Langevin equation, even
though $\rho({\bf x},t)$ is a stochastic field. Using
Eqs.~\ref{deterministic}, \ref{det2} and \ref{p3} it is possible to
write the mean field equation for the activity field:
\begin{equation}
\partial_t \rho({\bf x},t)=m(t)\rho({\bf x},t) 
- \rho({\bf x},t)\int_0^t dt' (p(t)+p(t')) \rho({\bf x},t')\,,
\label{det3}
\end{equation}
where $m(t)=p(t)s({\bf x},0)-c({\bf x},0)$.  In order to obtain the
complete functional Langevin equation (i.e., the field theory) for the
process in the neighborhood of the critical point (i.e., $p(t)\simeq
p_c$), we have to add to Eq.~\ref{det3} the noise term and the spatial
coupling terms.  The former is found simply by observing that, if in a
local region there are $k$ etchable sites in contact with the solution
at time $t$, an average number $p(t)k$ of them will be etched with a
typical Poissonian fluctuation of the order of $\sqrt{p(t)M}$.  This
shows that the noise term is $\sim \sqrt{\rho({\bf x}, t)}\eta ({\bf
x}, t)$ (multiplicative noise), where $\eta ({\bf x}, t)$ is the
typical white uncorrelated noise with zero mean and $\left<\eta ({\bf
x}, t)\eta ({\bf x'}, t')\right>=A\delta({\bf x}-{\bf
x'})\delta(t-t')$.  As a consequence of Eq.~\ref{det3} and of the form
of the noise term, it is possible to show that the only relevant term
of spatial coupling in the neighborhood of the critical point is the
diffusion term $\nabla^2\rho({\bf x}, t)$.  Since $p(t)$ is
smooth, positive and, in the neighborhood of the critical point,
$p(t)\simeq p_c>0$, at the end we obtain a field theory that belongs
to the universality class of {\em dynamical percolation} (i.e., of a
dynamical version of the classical percolation) \cite{dyn-perc}:
\begin{equation}
\partial_t \rho({\bf x},t)\!=\!m\rho({\bf x},t)\! -\! \gamma\rho({\bf
  x},t)\!\int_0^t dt'\rho({\bf x},t')\!+\!\nabla^2\rho({\bf x}, t)\!+\!
\sqrt{\rho({\bf x}, t)}\eta ({\bf x}, t)\,,
\label{det4}
\end{equation}
where $\gamma>0$, and $m$ measures, in the dynamical percolation ,the
distance $p-p_c$ between the chosen constant occupation probability
$p$ and its critical value $p_c$. If $m>0$ ---{\em active phase}--- the
process generate an infinite cluster of occupied sites, and if $m<0$
---{\em absorbing state} \cite{absorbing}--- the occupation process
arrests exponentially fast (i.e., $\rho({\bf x},t)$ decreases rapidly
to zero) leaving only a finite cluster of occupied sites. Since
$\rho({\bf x},t)\equiv 0$ is a solution of Eq.~\ref{det4}, once this
state is reached the occupation dynamics (i.e., in our model the
corrosion) stops spontaneously: $\rho({\bf x}, t)=0$ is a so-called an
{\em absorbing state}, and at $m=0$ we have an ``absorbing state phase
transition''. However, in our case (Eq.~\ref{det3}) $m$ depend on
time, and starting with a positive value, as $p(t)$ decreases in time,
it passes spontaneously from the active phase ($m>0$) to the absorbing
phase ($m<0$) arresting rapidly after this crossover. The velocity of
this crossover, that is, the duration of the critical regime, depends
on the finite size of the system which in this way determine both
the effective spatial gradient of $p$ and the
typical scale of fractality of the system, while the critical
exponents are those of percolation theory \cite{Munoz}. This shows,
finally, that our model is a self-organized version of Gradient
Percolation in any dimension.

\end{document}